\documentclass[twocolumn,aps,prl,superscriptaddress,showpacs,secnumroman,showkeys]{revtex4}
\usepackage{amsmath,bm}%
\usepackage{graphicx}%
\usepackage{epsf,epsfig,latexsym}
\usepackage{enumerate}
\usepackage{floatrow}
\usepackage{xcolor}

\usepackage{ulem} 

\begin{document}
\title{Evidence for high excitation energy resonances in the 7 alpha disassembly of $^{28}$Si  

}  
\author{X. G. Cao}
\affiliation{Shanghai Institute of Applied Physics, Chinese Academy of Sciences, Shanghai 201800, China}
\affiliation{Cyclotron Institute, Texas A$\&$M University, College Station, 
Texas 77843}
\author{E. J. Kim}
\affiliation{Cyclotron Institute, Texas A$\&$M University, College Station, 
Texas 77843}
\affiliation{Division of Science Education, Chonbuk National University, 567 
Baekje-daero Deokjin-gu,Jeonju 54896, Korea}
\author{K. Schmidt}
\affiliation{Institute of Physics, University of Silesia, 40-007 Katowice, Poland.}
\affiliation{Cyclotron Institute, Texas A$\&$M University, College Station, 
Texas 77843}
\author{K. Hagel}
\affiliation{Cyclotron Institute, Texas A$\&$M University, College Station, 
Texas 77843}
\author{M. Barbui}
\affiliation{Cyclotron Institute, Texas A$\&$M University, College Station, 
Texas 77843}
\author{J. Gauthier}
\affiliation{Cyclotron Institute, Texas A$\&$M University, College Station, 
Texas 77843}
\author{S. Wuenschel}
\affiliation{Cyclotron Institute, Texas A$\&$M University, College Station, 
Texas 77843}
\author{G. Giuliani}
\affiliation{Cyclotron Institute, Texas A$\&$M University, College Station, 
Texas 77843}
\affiliation{Laboratori Nazionali del Sud, INFN, via Santa Sofia, 62, 95123 Catania, Italy}
\author{M. R. D. Rodriguez}
\affiliation{Cyclotron Institute, Texas A$\&$M University, College Station, 
Texas 77843}
\affiliation{Instituto de F\'isica, Universidade de S\~ao Paulo, Caixa Postal 66318, CEP 05389-970, S\~ao Paulo, SP, Brazil}
\author{S. Kowalski}
\affiliation{Institute of Physics, University of Silesia, 40-007 Katowice, Poland.}
\author{H. Zheng}
\affiliation{Cyclotron Institute, Texas A$\&$M University, College Station, 
Texas 77843}
\affiliation{Laboratori Nazionali del Sud, INFN, via Santa Sofia, 62, 95123 Catania, Italy}
\author{M. Huang}
\affiliation{Cyclotron Institute, Texas A$\&$M University, College Station, 
Texas 77843}
\affiliation{College of Physics and Electronics Information, Inner Mongolia 
University for Nationalities, Tongliao, 028000, China}
\author{A. Bonasera}
\affiliation{Cyclotron Institute, Texas A$\&$M University, College Station, 
Texas 77843}
\affiliation{Laboratori Nazionali del Sud, INFN, via Santa Sofia, 62, 95123 Catania, Italy}
\author{R. Wada}
\affiliation{Cyclotron Institute, Texas A$\&$M University, College Station, 
Texas 77843}
\author{G. Q. Zhang}
\affiliation{Shanghai Institute of Applied Physics, Chinese Academy of Sciences, Shanghai 201800, China}
\affiliation{Cyclotron Institute, Texas A$\&$M University, College Station, 
Texas 77843}
\author{C.~Y. Wong}
\affiliation{Physics Division, Oak Ridge National Laboratory, Oak Ridge, USA}
\author{A. Staszczak}
\affiliation{Institute of Physics, Maria Curie-Sk{\l}odowska University, Lublin, Poland}
\author{Z. X. Ren}
\affiliation{State Key Laboratory of Nuclear Physics and Technology, School of Physics, Peking University, Beijing 100871, China}
\author{Y. K. Wang}
\affiliation{State Key Laboratory of Nuclear Physics and Technology, School of Physics, Peking University, Beijing 100871, China}
\author{S. Q. Zhang}
\affiliation{State Key Laboratory of Nuclear Physics and Technology, School of Physics, Peking University, Beijing 100871, China}
\author{J. Meng}
\affiliation{State Key Laboratory of Nuclear Physics and Technology, School of Physics, Peking University, Beijing 100871, China}
\affiliation{Yukawa Institute for Theoretical Physics, Kyoto University, Kyoto 606-8502, Japan}
\author{J. B. Natowitz}
\affiliation{Cyclotron Institute, Texas A$\&$M University, College Station, 
Texas 77843}

\date{\today}

\begin{abstract}
The excitation function for the 7 alpha de-excitation of $^{28}$Si nuclei 
excited to high excitation energies in the collisions of 35 MeV/nucleon 
$^{28}$Si with $^{12}$C reveals resonance structures that may indicate 
the population of toroidal high-spin isomers such as those predicted by 
a number of recent theoretical calculations.  This interpretation is 
supported by extended theoretical analyses.
\end{abstract}



\maketitle
 
\section*{I. INTRODUCTION}
Light nuclei in their ground states in the valley of stability usually have 
spherical or near-spherical geometries~\cite{stone05}. With increasing 
excitation energy and/or angular momentum, these nuclei may exhibit more 
exotic shapes~\cite{cohen74,wheeler61,brown80,aberg90}.  Theoretical 
investigations of the possible existence of extremely exotic nuclear shapes 
have a long history.  Wheeler suggested that, under certain conditions, 
nuclei could assume toroidal shapes~\cite{wheeler61}.  Pursuing this 
suggestion, Wong explored possible toroidal and bubble nuclei and predicted 
exited toroidal states in the mass region of $40\lesssim A\lesssim 70$ and 
$A$ $\lesssim$ 250 \cite{wong72,wong73,wong78}.  Large shell effects in light
nuclei and large Coulomb energies in heavy nuclei favor exotic toroidal 
configurations.  A recent search for heavy toroidal systems indicated that 
the probability of planar fragmentation configurations in the experimental 
data was much greater than predicted by quantum molecular dynamics 
calculations~\cite{najman15}. 

Extending such studies Wong and collaborator predicted that, toroidal 
configurations were also possible for nuclei with a sufficiently high 
angular momenta~\cite{wong78_1,wong85,zhang86}.
They defined the region of mass and angular momentum  in which such
configurations might be realized. More recent theoretical studies have
used microscopic techniques to address this question of light  toroidal
nuclei. In particular, Zhang $et~al.$ \cite{zhang10},
Ichikawa $et ~al.$ \cite{ichikawa12,ichikawa14} and Staszczak and Wong
\cite{staszczak14,staszczak15,staszczak15_1}, using varied approaches, 
have predicted the existence of toroidal isomers in light nuclei.   

We can understand the origin of possible light-A toroidal isomers in the 
following simple way.   In a nucleus with a toroidal shape, there are 
toroidal magic numbers $2(2m+1)$, with integer $m$$\ge$1, arising from 
large energy gaps between single-particle levels in the light mass 
region  (Fig.\ 1 of \cite{wong72}).  Extra stability  \cite{brack72} 
associated with toroidal  magic numbers leads to an excited  local energy 
minimum that is stable against the expansion and contraction of the 
toroidal major radius (see Fig.\ 2 of \cite{wong72}, Fig.\ 18 
of \cite{wong73}, and Fig.\ 1 of \cite{zhang10}).  Such an excited  
state residing in a local energy minimum under a  toroidal shape constraint 
will be called a diabatic state, and the corresponding constrained 
calculation a diabatic calculation \cite{zhang10}.  This is in contrast to 
an adiabatic state of the lowest energy minimum in an adiabatic calculation 
without a shape 
constraint~\cite{ichikawa12,ichikawa14,staszczak14,staszczak15,staszczak15_1}. 
Relative to a diabatic toroidal local energy minimum core, Bohr-Mottelson 
spin-aligning particle-hole excitations ~\cite{Boh81} can be constructed to 
yield a toroidal nucleus with a spin, $I$=$I_z$, by promoting nucleons 
with angular momentum aligned opposite to a chosen symmetry axis to populate 
orbitals with angular momentum aligned along the symmetry 
axis~\cite{ichikawa12,ichikawa14,staszczak14,staszczak15,staszczak15_1}.
The spinning toroidal nucleus possesses an effective 
``rotational'' energy that tends to expand the toroid, whereas the energy 
associated with the nuclear bulk properties tends to contract the 
toroid.  The balance between  the two energies gives rise to a local 
toroidal energy minimum \cite{wong78}.
For small values of $I$, the toroidal minimum occurs as an excited (diabatic) state above the sphere-like 
ground states.  As $I$ increases, the crossing of the toroidal and sphere-like energy surfaces takes place, and the toroidal high-spin energy minimum  switches to become the lowest energy (adiabatic) state.  Adiabatic self-consistent 
calculations have located an 
extensive region of toroidal high-spin isomer  (THSI) states that are stable against the 
expansion and contraction of the 
toroids~\cite{ichikawa12,ichikawa14,staszczak14,staszczak15,staszczak15_1}.

Motivated by the predictions of toroidal isomeric states in a number of 
light nuclei, we have undertaken searches for evidence of their existence.  
For $^{28}$Si, the nucleus investigated in the present work, the Staszczak 
and Wong calculations indicate that toroidal shapes with $I$= 0 become 
possible at excitation energies greater than 85 MeV. The existence of a 
stabilized state with angular momentum of $44\hbar$ and excitation energy 
of 143.18 MeV~\cite{staszczak14} was predicted. See Table 1.  

\begin{table}[t]
\begin{tabular}{|c|c|c|c|c|c|c|c|c|}
\hline
 &$I$       & $Q_{20}$ & $\hbar\omega$ & $E_x$ &$R$ & $d$ & $R/d$ & $\rho_{\rm max}$ \\
 & [$\hbar$]  & [b]      & [MeV]         &[MeV]& [fm]& [fm] &     & [fm$^{-3}$]   \\
\hline
$^{28}$Si & 44& -5.86    & 2.8           & 143.18& 4.33  & 1.45 & 2.99 & 0.119 \\
\hline

\end{tabular}
\caption{Calculated parameters for the predicted toroidal isomer in $^{28}$Si. Left to right: Spin $I$=$I_z$, quadrupole moment $Q_{20}$,  the cranking rotation frequency $\hbar \omega$, excitation energy $E_x$, toroidal major radius $R$, minor radius $d$, aspect ratio $R/d$, and maximum density $\rho_{\rm max}$}
\end{table}

In the calculation the toroidal shape of this short-lived isomer $^{28}$Si 
is characterized by a radius of 4.33 fm and a cross sectional radius of 
1.45 fm for the cylindrical ring containing the nucleons.  Thus the aspect 
ratio is 2.99.  The predicted toroidal states, although expected at very 
high energies, are analogous to yrast traps already observed in more 
conventionally shaped nuclei~\cite{ploszajczak78}.  Should such very 
highly excited stabilized toroidal states of light nuclei exist, their 
lifetimes should be short. They may de-excite or undergo shape relaxation 
rather quickly. In either case the most-likely de-excitation modes are 
particle or cluster emission and fragmentation~\cite{tsang06}.

It is well documented that macroscopic toroids fragment as a result of 
the development of Plateau-Rayleigh 
instabilities~\cite{plateau63,rayleigh14,pairam09,nurse15,fontelos11}.  In 
the basic Rayleigh description~\cite{rayleigh14} the dominance of a single
 mode of symmetric fragmentation leads to disassembly into equal size 
pieces, the number of which is of the order of the aspect ratio of the 
toroid~\cite{plateau63,rayleigh14,pairam09,nurse15,fontelos11}.  This 
correspondence is most accurate for large aspect ratios~\cite{nurse15}.  
Modern numerical simulations taking into account the viscosity of the 
fluid and the surrounding medium indicate that more complicated symmetric 
breakups involving different size fragments are possible~\cite{fontelos11}.  

As has been already discussed in the literature, nuclear tori might also 
manifest Plateau-Rayleigh instabilities~\cite{wong72,wong73,wong85}.  The 
aspect ratio predicted by Staszczak and Wong for the 143.18 MeV, $44\hbar$ 
state in $^{28}$Si suggests that, while other fragmentations are 
possible~\cite{nurse15},  the dominant instability of that toroid would 
lead to a break-up into $\sim 3$ fragments. However the actual dominant 
mode will be affected by the temperature dependent viscosity of the 
disassembling nucleus~\cite{hofmann08}.  In the nuclear case the 
discreteness of the nucleons, the existence of Coulomb forces, shell 
effects and variations in the fragment binding energies may also modify 
the fragmentation pattern of the torus. 

While strongly reduced Coulomb energies might be expected to provide a 
signature for the disassembly of predicted heavy or super-heavy nuclear 
toroids in their meta-stable ground-states~\cite{wong78,najman15}, this 
is probably not so for the predicted high spin light tori. These light 
tori are predicted to have very high excitation energies. Thus, in their 
disassembly, a lowering of the Coulomb repulsion between fragments is 
probably not observable as the release of large deformation and rotational 
energies will normally lead to large kinetic energies for the observed 
decay fragments. Finally, given the high excitation energies involved, 
it is likely that the initial fragments will often be excited and undergo 
subsequent de-excitations, smearing the signature of a Plateau-Rayleigh 
instability. 

All of these considerations suggest that judicious choices of reaction 
mechanism, exit channels and observables will be necessary to probe the 
possible existence of these very exotic and very interesting nuclei. 

Two notable features of the Staszczak-Wong and Ichikawa calculations are that
\begin{enumerate}
\item	the cross-sectional radii of the cylindrical rings  containing the 
nucleons are $\sim 1.5$ fm, essentially equal to the alpha particle radius. 
Indeed, in their search for isomeric states in $^{40}$Ca, 
Ichikawa \textit{et al.} used a ring of 10 alpha particles as an initial 
configuration for their cranked Hartree Fock calculations~\cite{ichikawa12}. 

\item	 the toroidal rings corresponding to the predicted isomers have 
matter densities approximately 2/3 of $\rho_0$ where $\rho_0$ is the normal 
central density of such nuclei~\cite{ichikawa14,staszczak14,staszczak15} .
\end{enumerate}

Since the general importance of $\alpha$-like correlations in the structure 
and properties of light nuclear systems at normal and reduced densities is now 
well established~\cite{beck14,ikeda68,oertzen06,freer07,kanada01,funaki08}, 
the features noted here suggest that the disassembly into alpha particles or 
alpha-conjugate nuclei might be favored for light nuclear toroids.  In 1986 
Wilkinson specifically suggested that spinning rings of alpha particles might 
be stabilized by circumnavigating neutrons. He discussed their stabilities 
toward electromagnetic or fragmenting de-excitations and the possibility of 
producing them in heavy ion collisions~\cite{wilkinson86}. The idea of 
stabilization by surrounding neutrons is also a feature of the extended 
Ikeda diagram systematics~\cite{oertzen14}.

We have searched for toroidal isomers in the disassembly of
$^{28}$Si produced in near-fermi-energy collisions of  $^{28}$Si with  
$^{12}$C. While some nucleon transfer and nucleon-nucleon collisions leading 
to early (pre-equilibrium)particle emission 
occur~\cite{hagel94,larochelle96,ono99,cassing87,charity10,mancusi10,lacroix04}, 
such collisions, many of which lead to essentially binary exit channels, 
are capable of producing projectile like nuclei in the A=28 mass region 
with high excitation energy and high angular momentum.  
Calculations~\cite{ono99,cassing87,charity10,mancusi10,lacroix04}indicate 
that an angular momentum in the range of 40$\hbar$  and an excitation energy 
as high as 170 MeV can be reached. These calculations should be considered 
as only indicative of possible angular momentum range as they do not have 
the ingredients to explore detailed quantum structure at such high 
excitation and angular momentum. When the excitation energy $E$ and the 
angular momentum $I$ of an emerging $^{28}$Si$^*$ corresponds to that of a 
toroidal high-spin isomer, the collective cranking motion and the 
rearrangement of the single-particle motion of the nucleons may eventually 
lead to the toroidal  high-spin isomer. 
 
Here we report results for an investigation of $^{28}$Si, focusing on 
the $7\alpha$ decay channels of excited projectile-like fragments produced 
in the reaction 35 MeV/nucleon $^{28}$Si + $^{12}$C.  In this reaction the 
energy available in the center of mass is 294 MeV, the maximum angular 
momentum, L$_{max}$, is $94\hbar$ (a reaction cross section of 2417 mb) , 
L$_{crit}$ for fusion is $26\hbar$ and the rotating liquid drop limiting 
angular momentum is $40\hbar$~\cite{wilcke80}.  These parameters indicate 
that the bulk of the reaction cross section will lead, not to fusion, but to 
initially binary configurations of excited projectile-like and target-like 
nuclei.  This is consistent with experimental results reported for similar 
collisions in this energy region~\cite{hagel94,larochelle96}.    

\section{Experimental Details}
The experiment was performed at Texas A$\&$M University Cyclotron Institute. A $35$ MeV/nucleon $^{28}Si$ beam produced by the $K500$ superconducting cyclotron  impinged on a $^{12}$C target. The reaction products were measured using a $4\pi$ array, NIMROD-ISiS (Neutron Ion Multidetector for Reaction Oriented Dynamics with the Indiana Silicon Sphere)~\cite{wuenschel09,wada04} which consisted of $14$ concentric rings covering from $3.6^{\circ}$ to $167^{\circ}$ in the laboratory frame~\cite{wuenschel09}. In the forward rings with $\theta_{lab}\leq 45^{\circ}$, two special modules were set having two Si detectors ($150$ and $500$ $\mu m$) in front of a CsI(Tl) detector ($3-10\ cm$), referred to as super-telescopes. The other modules (called telescopes) in the forward and backward rings had one Si detector (one of $150$, $300$ or $500$ $\mu m$) followed by a CsI(Tl) detector. The pulse shape discrimination method was employed to identify the light charged particles with $Z\leq3$ in the CsI(Tl) detectors. Intermediate mass fragments (IMFs), were identified with the telescopes and super-telescopes using the "$\Delta E-E$" method. In the forward rings an isotopic resolution up to $Z=12$ and an elemental identification up to $Z=20$ were achieved. In the backward rings only $Z=1-2$ particles were identified, because of the detector energy thresholds. In addition, the Neutron Ball surrounding the NIMROD-ISiS charged particle array provided information on average neutron multiplicities for different selected event classes.  Further details on the detection system, energy calibrations, and neutron ball efficiency can be found in~\cite{hagel94,wada04,lin14}. 

It is important to note that, for symmetric collisions in this energy range, the increasing thresholds with increasing laboratory angle lead to a condition in which the efficiencies strongly favor detection of projectile like fragments from mid- peripheral events. Modeling of these collisions using an Antisymmetrized Molecular Dynamics (AMD) code~\cite{ono99} and applying the experimental filter demonstrates that this is primarily an effect of energy thresholds.  

Nucleons are emitted during the initial phase of the collision. They are often modeled as emission from a \textit{virtual} mid-rapidity source having a velocity close to that of the nucleon-nucleon collision frame~\cite{cassing87}. While these nucleons, ejected prior to equilibration of the remaining system, are not from the de-excitation of the primary exit channel products, some of them appear in the projectile velocity region.  They should not be included in the calorimetric determination of the \textit{thermalized} excitation energy.   

\section{Analysis}
For the $^{28}$Si+$^{12}$C reaction, a total of 17.5 million events were 
recorded and a significant proportion of events had significant alpha-like (AL)
mass emission(i.e. alpha particles or alpha-conjugate nuclei). About $3.19 \times 10^5$ events had AL=28. Of these 8200 detected 
events had 7 alpha particles. 

The AMD calculations indicate that for detected A=28 channels $\sim 90 \%$ 
of the nucleons originate from the projectile.  Some mixing is consistent with 
damped collisions. A very careful inspection of experimental invariant 
velocity plots for each reaction channel and each emitted species confirmed 
that, projectile like sources strongly were strongly dominant in the 
detected events, except for Z=1 particles.  These Z=1 particles showed 
some contribution from both target-like and mid-rapidity sources, the 
latter characteristic of pre-equilibrium emission.  The observed source 
velocities for the selected A=28 events decrease slowly with exit channel 
complexity but are always above the center of mass velocity.  Complete 
fusion is a rare process and target like fragments are generally not 
detected as indicated above.  Searching for 8 alpha events we found that 
their yield was a factor of 10 less than for 7 alpha events.  This indicates 
small possibility of contamination from such events.  All of these features 
were taken into account in the following analysis.

To characterize the source excitation energies 
involved we have used calorimetric techniques to determine the excitation 
energy, $E_x$, of the primary projectile-like fragments.  $E_x$, is normally 
defined as the sum, of the kinetic energies of ejected particles and 
fragments in the frame of the total projectile-like nucleus, (determined 
by reconstruction of the mass and velocity of the primary excited nucleus 
from its de-excitation products) minus the reaction Q-value. See equation (1).
\begin{equation}
E_x = \Sigma_{i=1}^{M_{cp}}K_{cp}(i) + M_n<K_n> - Q
\end{equation}
where $M_{cp}$ is the multiplicity of charged particles, $K_{cp}$ is the 
kinetic energy of a charged particle in the source frame, $M_n$ is the 
neutron multiplicity and $<K_n>$ is the average neutron kinetic energy in 
the source frame.

\begin{figure*}
\epsfig{file=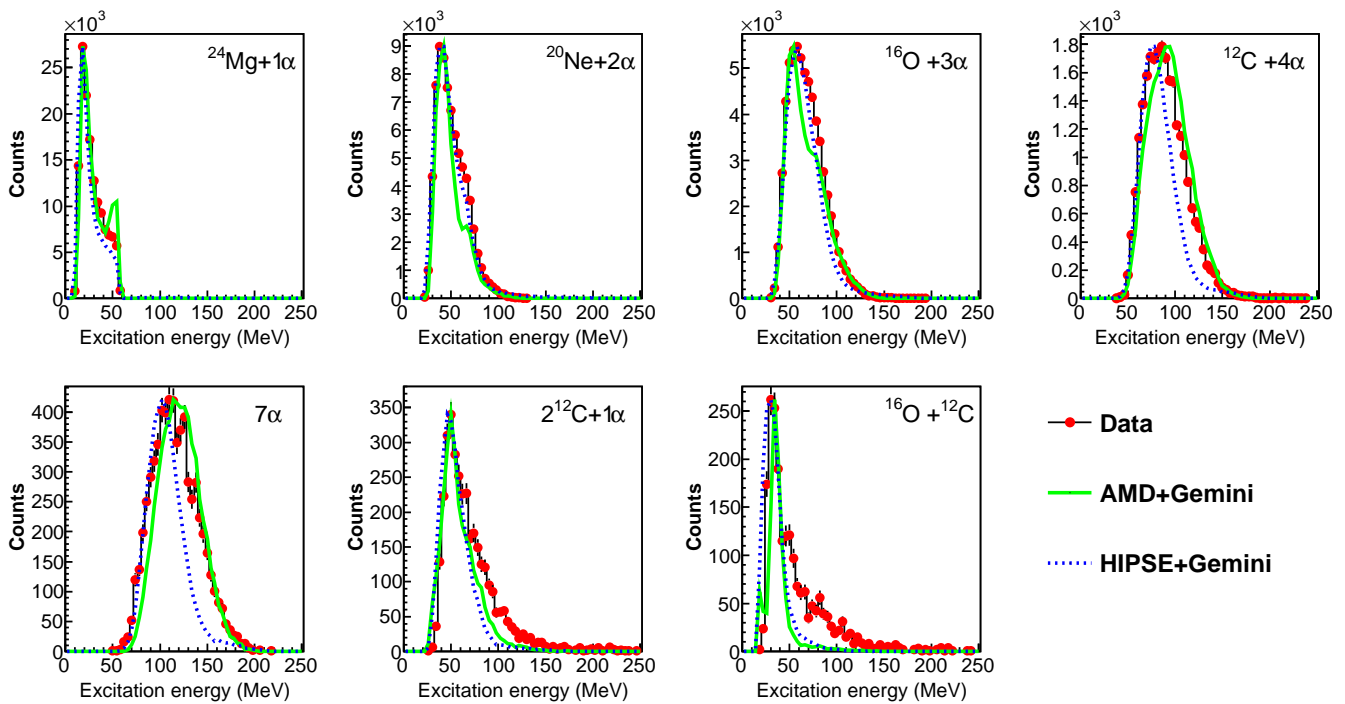,width=18cm,angle=0}
 \caption{Excitation functions for the alpha conjugate exit channels in the de-excitation of $^{28}$Si. The shapes of the experimental data are  compared with results of both AMD and HIPSE calculations. }
\label{fig1}
\end{figure*}

The events initially selected typically had a few Z=1 particles (and neutrons) and, in very rare cases, a heavier fragment, associated with them. In our event selection, which focusses on the alpha-conjugate exit channels, we have allowed Z=1 particles and neutrons in the events but we have determined the excitation energies excluding the energies of Z= 1 particles and neutrons as they are primarily pre-equilibrium particles, representing energy dissipation but not energy deposition into the projectile-like fragment~\cite{cassing87}. Invariant velocity plots for the alpha particles indicated that the alpha particles resulting from mid-rapidity pre-equilibrium emission were negligible and that the small number detected from the target like source (given the thresholds and geometry of the NIMROD detector) could be effectively removed by rejection of alpha particles with energies greater than 40 MeV in the projectile-like fragment  source frame.

In Figure~\ref{fig1} we present the excitation functions derived in this manner for the alpha-conjugate exit channels. For comparison, results from calculations using the phenomenological event generator HIPSE ~\cite{lacroix04} and the AMD transport model with a Gemini afterburner are also presented~\cite{ono99,charity10}. While the agreement between the data, the AMD and the HIPSE results is generally   good, we note that the 7-alpha exit channel results differ considerably. The experimental 7-alpha results appear to have structure at the higher excitation energies. In the following we focus on this exit channel.

Two hybrid codes, AMD-GEMINI~\cite{ono99,charity10,mancusi10} and HIPSE-GEMINI~\cite{charity10,mancusi10,lacroix04} were also used to calculate the 7-alpha excitation energy spectrum. The modeling of light particle emission using the code GEMINI was thoroughly examined in reference~\cite{mancusi10}.  In that work, formulations of the barriers and transmission coefficients, the level density and the yrast line were carefully explored. We have used the default parameter prescriptions derived from this work. Results of both calculations, filtered through the NIMROD acceptance and normalized to the data, are also presented in the figure. It is not expected that high energy resonances could appear in these models and, indeed, both are structure-less. The AMD results are somewhat broader in energy and shifted to slightly higher energy than the HIPSE results. 
 
As will be seen in this figure, the 7 alpha distribution spans the energy region in which toroidal configurations are predicted and the 143.18 MeV stabilized state is predicted to exist.  After reaching a maximum at $\sim 110$ MeV the excitation energy distribution shows some structure at 126 and 138 MeV.  The granularity and angular resolution of NIMROD-ISiS are not ideally suited to searches such as this, as the transformation to the source frame relies upon the angle of detection. Through simulations we have determined that the observed experimental width in excitation energy of an initially sharp state at 140 MeV will have a standard deviation, $\sigma$, of $\sim 4$ MeV resulting from the angular uncertainty. Taking this into account, the broad structures apparent in the excitation energy spectrum are consistent with much narrower resonances in the excitation energy distribution. We have checked this by adding a 7 alphas delta function with Ex=143 MeV to the uncorrelated 7 alpha events and observing that the resultant filtered spectrum is consistent with our observed spectrum. Clearly an experiment with much better angular resolution, allowing better resolution for the excitation energy spectrum, would be very desirable. 

\begin{figure}
\epsfig{file=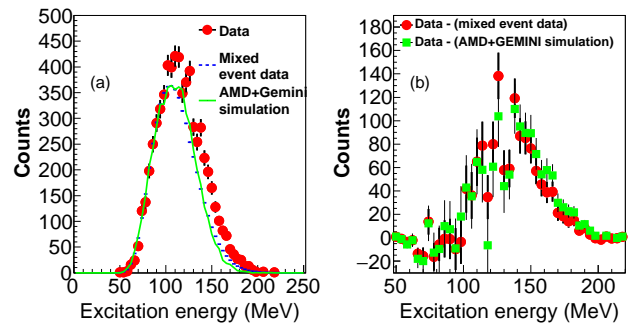,width=8.5cm,angle=0}
 \caption{Excitation energy distribution leading to observed 7-alpha events. 
Left panel-the data are represented by a solid black line. An uncorrelated 
spectrum derived from event mixing is represented by a long dashed line. The 
filtered result from an AMD-GEMINI calculation is indicated by the dashed and 
dotted line (see text).  The last two are normalized to the experimental 
spectrum at the lower edge of the spectrum. On the right the differences 
between the experimental spectrum and the others are presented.  Relative 
to the uncorrelated background derived from the experiment the statistical 
significance of the difference peak at 114 MeV is $5.3\sigma$, 
at 126 MeV is $8.0\sigma$ and at 138 MeV is $7.2\sigma$. See text}
\label{fig2}
\end{figure}

In figure~\ref{fig2}, on the left hand side, we compare the experimental spectrum for the 7 alpha events to an uncorrelated 7 alpha spectrum. The spectrum of uncorrelated events was constructed by randomly selecting 7 alpha particles from 7 different events. The random selection was done many times to assure that the statistical fluctuations for this spectrum would be much lower than those of the correlated event spectrum. This uncorrelated spectrum is taken to represent the 7 alpha phase space in this excitation energy region. Subtracting the normalized uncorrelated spectrum from the data results in the difference spectrum depicted in the right of Figure~\ref{fig2}. Some excess is seen at 114 MeV. The peaks at 126 and 138 MeV are quite prominent and there is a tailing toward higher energies. The under-shoot at 100 MeV and below may suggest that the normalization of the uncorrelated spectrum is too conservative. If that spectrum is lowered the peaks in the difference spectrum will be even more prominent. For comparison we have also employed the AMD calculated spectrum as a background spectrum to be subtracted from the experimental spectrum. For this purpose the AMD-GEMINI spectrum was shifted to agree with the experimental uncorrelated spectrum at the lower edge of the experimental spectrum. 

The difference spectrum obtained by subtracting the normalized AMD- GEMINI spectrum from the experimental data, is very similar to, but not exactly the same as that obtained when subtracting the experimental uncorrelated spectrum from the experimental data.  

The structure in the experimental $7\alpha$ spectrum appears to reflect 
mechanisms not encompassed in the dynamic reaction models or normal 
statistical decay treatments.  Relative to the uncorrelated background 
derived from the experiment the statistical significance of the difference 
peak at 114 MeV is $5.3\sigma$, at 126 MeV is $8.0\sigma$ and at 
138 MeV is $7.2\sigma$~\cite{bityukov98}.  We take these to be the minimum 
values for the statistical significance since the construction of the 
uncorrelated spectrum includes contributions from the peak region and the 
number of uncorrelated alphas may, therefore, be overestimated in that region.

Evaluation of the statistical significance of the observed peaks is sensitive 
to the background assumed. If we base the test for the statistical 
significance on the use of the AMD GEMINI result we find a statistical 
significance of 114, 126 and 138 MeV  peaks  to be 
$4.2\sigma$, $6.0\sigma$ , and $6.6\sigma$, respectively.

\section{Cross Section and Angular Momentum}
To determine the cross section for the 7 alpha channel we have assumed that 
the total events detected with our minimum bias trigger (1 particle or 1 
fragment detected), corrected for detector efficiency, represent a total 
reaction cross section of 2417 mb~\cite{wilcke80}. The overall detector 
efficiency was determined from the ratio of numbers of AMD-GEMINI generated 
events before and after the detector filter.  The specific efficiency for 
the 7 alpha channel was also determined in a similar fashion.  Because double 
hit corrections can be large, these results are very sensitive to the number 
of, and excitation energies of intermediate $^8$Be nuclei produced.  Thus 
there may be a significant systematic uncertainty in the resultant cross 
section.  Based on the estimated uncertainties we estimate the cross section 
for the 7 alpha channel to be 1.9 mb with a systematic uncertainity of 
$\sim 25\%$.  We estimate the cross sections for the 126 and 138 MeV peaks,
respectively, to be $51 \mu$b and $28 \mu$b with similar uncertainities. 

If the state observed at 138 MeV corresponds to the predicted 143.18 MeV 
toroidal state the angular momentum would be $44\hbar$. Our results give 
no direct information on the angular momentum.  As already noted, 
the AMD and HIPSE calculations, employing semi classical techniques, indicate 
that angular momenta in the range of $40\hbar$ are reached but they do not 
have the ingredients necessary to explore detailed structure at such high 
excitation energy and angular momentum.  

\begin{figure*}
\epsfig{file=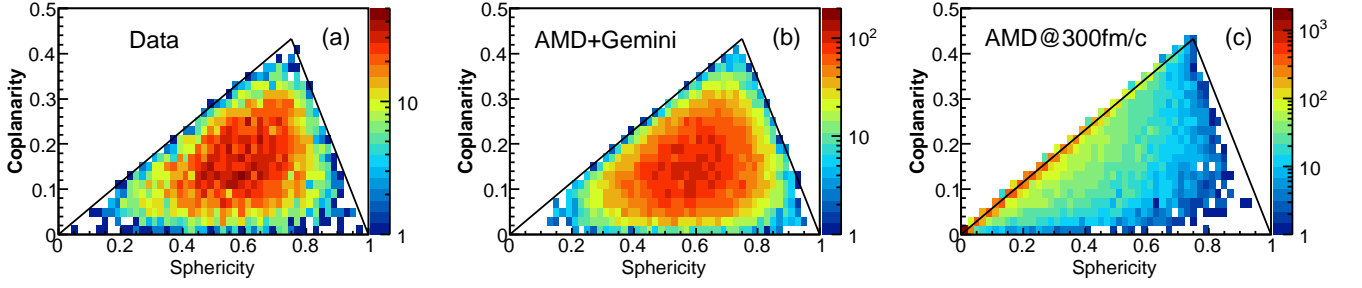,width=18cm,angle=0}
 \caption{Shape analysis of alpha-conjugate exit channels in the  de-excitation of  AL = 28,  Z = 14 projectile-like nuclei produced in the 35 MeV/nucleon $^{28}$Si + $^{12}$C reaction. Left-Experimental data.  Middle-Filtered results from AMD-GEMINI calculation.  Right- Results for AMD primary fragments at 300 fm/c. 
}
\label{fig3}
\end{figure*}
 
\section{Momentum Space Shape Analyses}
We have utilized a shape analysis technique~\cite{fai83,bondorf90} to diagnose the momentum space source shape for the $7\alpha$ events.  This type of analysis is a popular method to study emission patterns of sources, dynamical aspects of multi-fragmentation and collective flows of particles in heavy ion collisions at intermediate and relativistic energies. Although we are not able to observe directly the geometric shape of these sources, the momentum space correlations among the detected fragments provide hints of shapes and information on the disassembly dynamics.  Should nuclei with exotic shapes undergo simultaneous fragmentation into equal sized cold fragments the observed momentum space distributions of products would be directly related to the initial geometric configuration of the de-exciting system. In contrast, if the primary fragments are excited or the emission is sequential, de-excitation of the initially produced fragments could significantly modify the initial momentum space distribution.  For the light alpha-conjugate nuclei explored in this study, alpha emission from excited heavier primary fragments would produce large perturbations of that distribution.  

The analysis employs a tensor constructed from the product momenta, written as: $T_{ij}=\Sigma_{\nu=1}^N p_i^\nu  p_j^\nu$ where $N$ is the total nucleon number, $p_i^\nu$  is the momentum component of $\nu^{th}$ nucleon in the center-of-mass and i refers to the Cartesian coordinate. The tensor can be diagonalized to reduce the event shape to an ellipsoid. The eigenvalues of the tensor: $\lambda_1$, $\lambda_2$ and $\lambda_3$ , normalized by: $\lambda_1+\lambda_2+\lambda_3=1$ and ordered according to: $\lambda_1\leq\lambda_2\leq\lambda_3$, can quantitatively give  shape information of the events. The sphericity is defined as: $S= {{3}\over{2}}(1-\lambda_3)$, and coplanarity is defined as: $C= {\sqrt{3}\over{2}}(\lambda_2-\lambda_1)$. In the sphericity-coplanarity plane, the ideal rod, disk and sphere events are exactly located at the three vertices of the triangle: (0,0) (3/4, $\sqrt{3}$/4) and (1,0), respectively. Two fragment events will always appear at 0,0 while three fragments events will appear on the rod-disk axis. To appear as a disk at the apex of this triangle requires a simultaneous and symmetric fragmentation into 3 or more equally sized pieces.

The results of the momentum shape analysis are shown in Fig. 3(a).  While 
there are some events located in the disk region, the bulk of the events are 
elsewhere. Also depicted in Figure~\ref{fig3} are sphericity-coplanarity plots of the 
AMD-GEMINI results. Figure~\ref{fig3}(b) portrays the filtered final results.  
Figure~\ref{fig3}(c) presents the plot for the 300 fm/c freezeout momentum 
distribution predicted by AMD. Clearly this freeze-out distribution is much 
more rod to disk like than that observed after de-excitation and filtering. 
The model results suggest that the observed final distribution normally 
results from processes in which an initial breakup into larger excited 
fragments is followed by alpha particle de-excitation of those fragments.

\vspace*{0.2cm}

\section{Comparison with Toroidal Shell Model }

The 7$\alpha$  de-excitation of $^{28}$Si projectile-like fragments encompasses 
the excitation energy range for which toroidal configurations are predicted 
to exist. The resonance structure 
indicates the existence of 
stabilized configurations which may correspond to toroidal high-spin states such as 
those predicted in several theoretical 
calculations~\cite{zhang10,ichikawa12,ichikawa14,staszczak14,staszczak15,staszczak15_1}.   If the 
observed state at 138 MeV corresponds to the 143.18 MeV state predicted to 
exist by Staszczak and Wong \cite{staszczak14}, the angular momentum would be $44\hbar$.  

Resonances at such high excitation energies are rather unusual, and we know 
of no model except the toroidal isomer model that predicts $^{28}$Si$^*$ 
resonances at these high excitation energies. 
We  have explored whether the experimental
data may be described in terms of the toroidal shell model in which  nucleons move in the toroidal potential 
$V(\rho,z)=\frac{1}{2}m\omega_0^2(\rho -R)^2+\frac{1}{2}m\omega_0^2z^2$ \cite{wong72,wong73,staszczak14}.
Upon neglecting the small spin-orbit interaction,
the toroidal single-particle 
energy $\epsilon({n\Lambda\Omega})$  for the   $|n \Lambda \Omega\rangle$ 
 state, in the $I$=0  toroidal core with a major radius $R$ and $R\gg d$,   is approximately
\begin{eqnarray}
 \epsilon({n\Lambda\Omega}) \approx  \hbar \omega_0 (n+1) +\frac{\hbar^2 \Lambda^2}{2 m R^2},
\label{sp}
\end{eqnarray}
where $n$=$n_z$+$n_\rho$ is the harmonic oscillator quantum number, 
 $\omega_0$ the oscillator frequency,  $m$  the nucleon mass,
$\Lambda$ the orbital angular momentum about the symmetry $z$-axis, 
$\Omega_z$=$\Lambda$+$s_z$, and  $s_z$ the intrinsic nucleon spin.

Relative to the toroidal $I$=0 core at $E_0$ occupying the lowest toroidal single-particle states,
the spin-aligning Bohr-Mottelson particle-hole excitations leading to toroidal high-spin isomers  can be constructed  
 by following the crossings of Routhian energy levels as a function of increasing cranking frequency $\hbar \omega$  (Fig.\ 1(b) of
\cite{staszczak14}).
The contribution to $\Delta I_z$ and $\Delta(E_I-E_0)$ from a particle-hole  excitation can be easily obtained 
 from the  changes in particle-hole state quantum numbers $(n\Lambda\Omega)$ and $\epsilon(n\Lambda\Omega)$. 
For $^{28}$Si, for example,  the 1p-1h excitation 
contributes $\Delta I_z$=8$\hbar$, and $\Delta(E_I-E_0)$=7$\hbar^2/2 m R^2$
by promoting
a nucleon from $| 03(-7/2)\rangle$  to  
$|04(9/2)\rangle$.  Similar contributions can be obtained  for  the 2p-2h  excitation
 by additional  promotion from
$|03(-5/2)\rangle$  to  $|04(7/2))\rangle$,
3p-3h  from
$|02(-5/2)\rangle$  to  $|05(11/2)\rangle$, and  4p-4h 
 from
$|02(-3/2)\rangle$  to  
$|05(9/2)\rangle$.
From such calculations, we obtain 
the spin $I$=$I_z$  and the relative  energy $(E_I-E_0)$ 
(in units of $(\hbar^2/2 m R^2)$), for various  $^{28}$Si$^*$  toroidal high-spin 
isomer states as the signature for toroidal $^{28}$Si$^*$  in Table \ref{tableII}.

\begin{table}[h]
\caption { Toroidal  high-spin  isomers  (THSI) of  $^{28}$Si$^{^*}$ in the toroidal shell model.  The spin-aligning (n particle)-(n hole) excitations, for neutrons $(\nu)$ and protons $(\pi)$,
relative to a  toroidal core with $I$=0 and energy $E_0$,
  lead to the THSI state of spin $I$=$I_z$, and excitation energy $E_I$. }
\vspace*{0.3cm}
\begin{tabular}{|c|c|c|c|}
\hline
  Configurations      &     $I$  & $(E_I-E_0)$ &$E_I$\\
            & & in $\hbar^2/2mR^2$   & (MeV)\\
   \hline
(0p-0h)$_\nu$(0p-0h)$_\pi$  &  0 & 0   &91.82  \\
\hline
(1p-1h)$_\nu$(1p-1h)$_\pi$  &  16 &  14  &101.2   \\
\hline
(0p-0h)$_\nu$(2p-2h)$_\pi$+(2p-2h)$_\nu$(0p-0h)$_\pi$ &  14 &  14  &101.2   \\
\hline
(2p-2h)$_\nu$(2p-2h)$_\pi$  & 28  & 28  &110.58   \\
\hline
(2p-2h)$_\nu$(3p-3h)$_\pi$+(3p-3h)$_\nu$(2p-2h)$_\pi$ & 36 & 49&124.65    \\
\hline
(3p-3h)$_\nu$(3p-3h)$_\pi$ &  44 & 70& 138.72  \\
\hline
(3p-3h)$_\nu$(4p-4h)$_\pi$+(4p-4h)$_\nu$(3p-3h)$_\pi$  & 50 & 91 & 152.79\\
\hline
(4p-4h)$_\nu$(4p-4h)$_\pi$ & 56 & 112 &166.86\\
\hline
\end{tabular}
\label{tableII}
\end{table}

We can identify the $I$=44$\hbar$, 36$\hbar$ and 28$\hbar$ toroidal high-spin 
isomers in the toroidal shell model as the resonances $E_A$=138.7 MeV, 
$E_B$=125.4 MeV, and $E_C$=112.7 MeV,  respectively, with a high degree of 
confidence based on the following grounds: (i) the theoretically predicted 
$I$=44$\hbar$ state at $E_{44}=143$ MeV in Ref.~\cite{staszczak14} is close 
to the observed experimental resonance energy $E_A$=138.7 MeV, (ii) the 
theoretical energy  spacing between the $I$=44$\hbar$ an $I$=36$\hbar$ THSI 
states and between the $I$=36$\hbar$ and the $I$=28$\hbar$ THSI states are 
equal, and likewise the experimental spacing between resonances $E_A$ and 
$E_B$ and between resonances $E_B$  and $E_C$  are also approximate equal,
and (iii) above the $I$=44$\hbar$ THSI state,  there exist, theoretically, 
additional $I$=50$\hbar$ and 56$\hbar$ THSI states at higher energies, and 
there remains significant experimental cross section above the resonance 
$E_A$=138.7 MeV.  Expressing $E_I$ as  
\begin{eqnarray}
E_I= E_0 + [(E_I-E_0)/(\hbar^2/2 m R^2) ]  \times (\hbar^2/2 m R^2),
\label{EIeq}
\end{eqnarray}
we find  by using  the quantity $[[(E_I-E_0)/(\hbar^2/2 m R^2) ]$ in 
Table~\ref{tableII} that $E_0$= 91.82 {\rm MeV} and $R$=5.56 fm,  with 
which we can determine $E_I$ of  all other THSI states presented in 
Table~\ref{tableII}.    Table {\ref{tableII}}  reveals that up to 4p-4h 
excitations,  the toroidal shell model predicts  10 THSI yrast states built 
on the toroidal  $I$=0 core.  They occur within the range of excitation 
energies of the present measurement.  Note that some of the states are 
doubly degenerate.

Based upon the above THSI energy levels $E_I$ as primary ingredients, we 
have constructed a simple phenomenological semi-empirical formula to 
estimate the relative THSI production cross sections by including the width 
parameters  $\sigma_I$.  We start by assuming that the $7\alpha$ cross 
section obtained after subtracting the uncorrelated cross section from the 
correlated cross section in Fig.~\ref{fig2}(b) arises dominantly from 
toroidal configurations.  We assume further that the reaction cross section 
above the observed threshold is proportional to the distribution of deposited 
angular momentum $I$ which is governed by the impact parameter and the 
collision dynamics.
 
For toroidal THSI production, the energy and angular momentum must match 
that of a THSI state.  Hence the sum over $I$ with $(dI)$=1 is carried out 
for THSI states. Neglecting  other unknown  factors,  we developed 
the following semi-empirical cross section formula for the 7$\alpha$ 
channel from toroidal configurations 
\begin{eqnarray}
\sigma_{\rm toroidal}(E_x, 7\alpha)&=&A\sum_{I=I_{\rm toroid}} \frac{ g_{{}_I} I }{1+\exp\{ (I-I_{\rm max})/{a} \}}
\nonumber\\
&&\times  \frac{1}{\sqrt{2\pi}\sigma_I} \exp\{ - (E_x-E_I)^2/{2\sigma_I^2} \},~~~~~
\label{semiempirical} 
\end{eqnarray}
where gi is the state degeneracy factor and $I_{\rm max}$ and the diffusion 
parameter $a$ are introduced phenomenologically to describe initial-state 
dynamical   and/or final-state structural limitations. 

\begin{figure}
 \epsfig{file=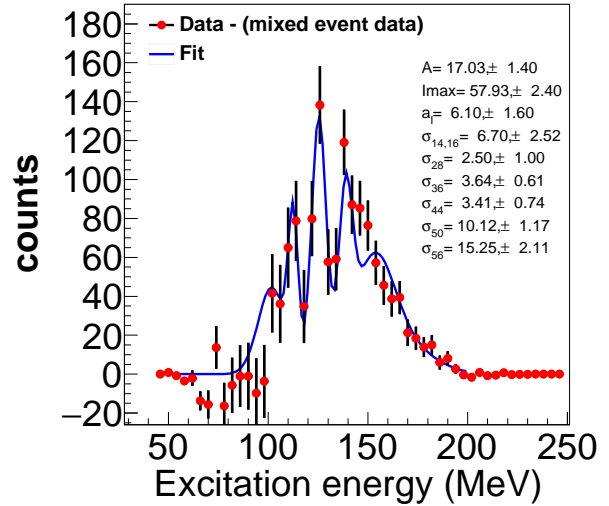,width=8cm,angle=0}
\caption{The  experimental (correlated  data) - (mixed event data)  
(in solid points) are compared with the results of semi-empirical 
formula (\ref{semiempirical}) containing the signature of the toroidal 
high-spin isomers as primary ingredients (in solid curve) with the list 
of extracted widths and other parameters.
}
\label{fig4}
\end{figure}

The most important primary ingredients in the above formula are the THSI 
spin $I$, energies $E_I$, and degeneracies  $g_{{}_I}$.  Upon using the 
toroidal shell model of Table~\ref{tableII}  to fix these primary 
ingredients, we extract  the secondary quantities of the widths and other 
parameters.  We find that the gross features of the excitation function 
can be well described by the semi-empirical formula~(\ref{semiempirical}) 
with extracted widths and fitting parameters as shown in 
Fig~\ref{fig4}.  The extracted widths for the sharp $I$=44$\hbar$,  
36$\hbar$ and 28$\hbar$ resonances are small, of the order of the 
experimental bin size.  The extracted widths for the $I$=50$\hbar$ 
and $56\hbar$ states are large, which  may indicate that the particle-hole 
excitations for these highest lying THSI states may involve promoting 
particles to populate states with large intrinsic widths close to the 
particle drip line.  However, low statistical uncertainities and 
uncertainities in background in that region may contribute to the apparent
broadening.

The present analysis indicates that the a spinning toroidal $^{28}$Si$^*$  
expresses itself as a set of THSI states with a unique signature listed  
in Column 3 of  Table \ref{tableII}.  In the excitation energy spectrum we 
find different facets of the toroidal signature,i.e, the presence of 
sharp resonances at appropriate energies, the spacing between some of the 
resonances, the apparent presence of  THSI states at energies  higher than 
the predicted $I$=44$\hbar$ state. 

The approximate matching of the experimental excitation energy spectrum 
with the semi-empirical cross section formula containing all THSI states up 
to 4p-4h excitations as exhibited in Fig.\ (\ref{fig4}), provides strong 
evidence for  the possible production of toroidal high-spin isomers in the 
present experiment.

\section{Additional Theoretical Support from Relativistic Mean-Field CDFT  Theory} 

Our experimental results and the phenomenological analysis suggest the 
existence of more than one stabilized state.  Previous non-relativistic 
adiabatic calculations without  a toroidal shape constraint give only  
the $I$=44$\hbar$ state \cite{staszczak14}, and cannot be used to calculate  
diabatic THSI states such as those with  $I $$< $44$\hbar$.  The 
phenomenological description of the toroidal shell  model predicts a total 
of 10 THIS states built on the toroidal $I$=0 core.  These toroidal isomers 
states can be searched for and examined by the covariant density functional 
theory (CDFT)~\cite{meng16}, which exploits basic properties of QCD at 
low energies, in particular, symmetries and the separation of scales.  It 
is worth pointing out that the CDFT theory has  provided an excellent 
description of ground states and excited (diabatic) states for nuclei all 
over the periodic table with a high predictive 
power~\cite{meng06,meng13,liang15}.  Using a universal density functional 
and without assuming the existence of clusters {\it a priori}, CDFT provides a 
high degree of confidence in the investigation of nuclear toroidal structures.

With the most successful density functionals PC-PK1~\cite{zhao10} and 
DD-ME2~\cite{lalazissis05}, the newly-developed cranking CDFT in 3D lattice 
space~\cite{ren17,ren18} has been applied to investigate the toroidal states 
in $^{28}$Si.  In these calculations, the $z$ axis is chosen as the symmetry 
axis. Grid points $34\times34\times24$ are respectively taken for 
$x$, $y$ and $z$ with a step size 0.8 fm.  The self-consistency of 
calculations is achieved with an accuracy of $10^{-4}$ MeV for the 
single-particle levels.  The pairing correlations are neglected.

By choosing a trial initial wave function with ring-like configuration of 
seven alpha-particles on the plane with $z=0$, a toroidal state with 
$I = 44\hbar$ is obtained at rotational frequency $\hbar\omega = 2.5$ MeV,  
corresponding to 3p-3h configurations for both neutrons and protons.  The 
excitation energy of this toroidal state is 147.93 MeV for PC-PK1 and 
145.08 MeV for DD-ME2.

\begin{figure}
  \epsfig{file=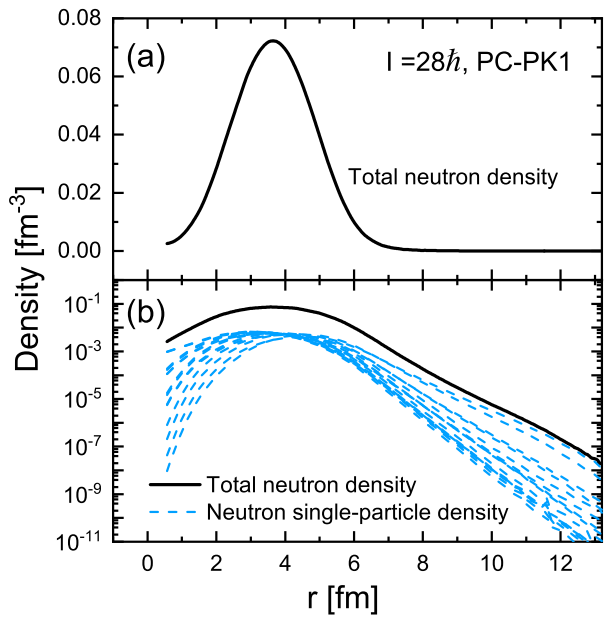,width=8cm,angle=0}
 \caption{Neutron radial density distributions ($z$ direction is integrated) 
  of the occupied single-particle levels (blue and thin lines)
  as well as the total density distribution (black and thick lines) in 
  toroidal state with $I=28\hbar$.
}
\label{fig5}
\end{figure}

Local energy minima are found for the toroidal state with $I$=28$\hbar$, 
corresponding to  2p-2h excitations for both neutrons and protons,
and the toroidal state with $I$=36$\hbar$, corresponding to the 2p-2h 
excitation for neutrons and 3p-3h for protons.  The toroidal state with 
$I = 44\hbar$ can be easily obtained by the adiabatic CDFT calculations.
For the toroidal states with $I = 28\hbar$ and 36$\hbar$, diabatic or 
configuration-fixed CDFT calculations are necessary.
Additional toroidal states with $I$=0$\hbar$, 14$\hbar$, 16$\hbar$, 
50$\hbar$ and 56$\hbar$ have also been  located.  They are shown in 
Table~\ref{Tab_property_torus_PC-PK1} and \ref{Tab_property_torus_DD-ME2}, 
respectively.  In the column of $\hbar\omega$, the mid-points of a range of 
the cranking rotational frequencies are given for toroidal diabatic (excited) 
state configurations arising from particle-hole excitations without 
specifying $\hbar \omega$.

\begin{table}
    \caption{Properties of toroidal states in $^{28}$Si obtained by the 
covariant functional PC-PK1~\cite{zhao10}.  In the table, $\hbar\omega$ 
is rotational frequency, $\beta_{20}$ is quadrupole deformation, related 
to the quadrupole moment $\langle Q_{20}\rangle$  of \cite{staszczak14}  
by $\beta_2$=$\langle Q_{20}\rangle / [(3/\sqrt{5\pi})A^{5/3} r_0^2]$ 
with $r_0$=1.2 fm, $E^*$ is excitation energy, and $R$, $d$, 
and $\rho_{\rm max}$ are determined by fitting the calculated density 
distributions to the Gaussian function 
$\rho(x,y,z)=\rho_{\rm max}e^{-[(\sqrt{x^2+y^2}-R)^2+z^2]/(d^2/\ln2)}$.
}\label{Tab_property_torus_PC-PK1}
\hspace*{-0.5cm}
   \begin{tabular}{lcccccccc}
\hline
 $I$&Configuration&$\hbar\omega$   &$\beta_{20}$  &$E^*$   &$R$  &$d$    &$\rho_{\textrm{max}}$\\
    $[\hbar]$         &                         &[MeV]           &              &[MeV] &[fm] &[fm]    &[fm$^{-3}$]\\
     \hline
     0         &(0p0h)$_\nu$(0p0h)$_\pi$     &---             &$-1.29$       &72.65   &3.37 &1.32      &0.167  \\
     14        &(0p0h)$_\nu$(2p2h)$_\pi$     &---             &$-1.42$       &91.04   &3.53 &1.34      &0.156  \\
     $14^*$    &(2p2h)$_\nu$(0p0h)$_\pi$     &---             &$-1.43$       &91.34   &3.54 &1.34      &0.156  \\
     16        &(1p1h)$_\nu$(1p1h)$_\pi$     &---             &$-1.42$       &89.56   &3.53 &1.33      &0.158  \\
     28        &(2p2h)$_\nu$(2p2h)$_\pi$     &  $\sim$2.49
&$-1.54$       &106.26  &3.68 &1.34      &0.149  \\
     36        &(2p2h)$_\nu$(3p3h)$_\pi$    & $\sim$2.58
&$-1.76$       &128.14  &3.96 &1.34     &0.137  \\
     $36^*$    &(3p3h)$_\nu$(2p2h)$_\pi$     &---             &$-1.77$       &128.45  &3.98 &1.34    &0.137  \\
     44        &(3p3h)$_\nu$(3p3h)$_\pi$     & $\sim$2.81
&$-2.02$       &147.92  &4.27 &1.34    &0.127  \\
     50        &(3p3h)$_\nu$(4p4h)$_\pi$     &---             &$-2.38$       &167.95  &4.68 &1.34      &0.116  \\
     $50^*$    &(4p4h)$_\nu$(3p3h)$_\pi$     &---             & $-2.39$       &168.18  &4.69 &1.34     &0.115  \\
     56        &(4p4h)$_\nu$(4p4h)$_\pi$     & $\sim$2.71
&$-2.79$       &185.18  &5.11 &1.34      &0.106  \\
\hline
   \end{tabular}
\end{table}

\begin{table}
    \caption{Same as Table. \ref{Tab_property_torus_PC-PK1} but for the covariant functional DD-ME2~\cite{lalazissis05}.
    }\label{Tab_property_torus_DD-ME2}
  \begin{ruledtabular}
   \begin{tabular}{lccccccc}
     $I/\hbar$ &Configuration            &$\hbar\omega$   &$\beta_{20}$  &$E^*$   &$R$  &$d$     &$\rho_{\textrm{max}}$\\
               &                         &[MeV]           &              &[MeV]   &[fm] &[fm]     &[fm$^{-3}$]\\
     \hline
     0         &(0p0h)$_\nu$(0p0h)$_\pi$     & $\sim$0.01
&$-1.24$       &65.31   &3.31 &1.24     &0.190  \\
     14        &(0p0h)$_\nu$(2p2h)$_\pi$     &---             &$-1.35$       &85.11   &3.43 &1.26    &0.177  \\
     $14^*$    &(2p2h)$_\nu$(0p0h)$_\pi$     &---             &$-1.36$       &85.53   &3.45 &1.26      &0.177  \\
     16        &(1p1h)$_\nu$(1p1h)$_\pi$     &---             &$-1.35$       &83.42   &3.44 &1.25      &0.179  \\
     28        &(2p2h)$_\nu$(2p2h)$_\pi$     & $\sim$2.70
&$-1.46$       &101.08  &3.58 &1.27      &0.168  \\
     36        &(2p2h)$_\nu$(3p3h)$_\pi$     & $\sim$2.76
 &$-1.68$       &124.32  &3.85 &1.28      &0.151  \\
     $36^*$    &(3p3h)$_\nu$(2p2h)$_\pi$     &---             &$-1.69$       &124.73  &3.87 &1.28   &0.151  \\
     44        &(3p3h)$_\nu$(3p3h)$_\pi$     &$\sim$2.88
&$-1.95$       &145.07  &4.18 &1.30     &0.136  \\
     50        &(3p3h)$_\nu$(4p4h)$_\pi$     &---             &$-2.35$       &165.85  &4.64 &1.32    &0.118  \\
     $50^*$    &(4p4h)$_\nu$(3p3h)$_\pi$     &---             &$-2.37$       &166.13  &4.66 &1.32   &0.118  \\
     56        &(4p4h)$_\nu$(4p4h)$_\pi$     & $\sim$2.64
&$-2.83$       &183.27  &5.14 &1.34    &0.104  \\
   \end{tabular}
  \end{ruledtabular}
\end{table}

In order to investigate the stability of these high-spin torus isomers 
against particle emission, we have examined the radial density distributions 
of the occupied single-particle levels as well as the total density 
distributions for these toroidal states.  As an example, neutron densities 
in toroidal state with $I=28\hbar$ are shown in figure 5.  It can be clearly 
seen that all radial density distributions are well localized and the 
stability of the toroidal isomer against particle emission is hereby 
demonstrated.  Similar conclusions can be drawn for other toroidal states as 
well.

Results from the CDFT theory confirm the previous theoretical result 
predicting a toroidal high-spin  isomer state with $I$=44$\hbar$ at an 
excitation energy of $E_x$ 145 - 148 MeV, and support the identification of 
the resonance observed experimentally at $E_x$=138 MeV as this possible 
toroidal state.  In addition all THSI states obtained in the toroidal shell 
model have also been located in the relativistic mean-field CDFT theory, 
supporting the use of the THSI states in the toroidal shell model as the 
signature for toroidal high-spin isomers.
 
There is however one notable difference that may reveal new physics 
associated with the toroidal shape.  The spacings between energy levels 
in the relativistic CDFT theory appear to be significantly greater than 
their corresponding spacings in the toroidal shell model, or in the 
experimental data. One interesting possibility may be that the toroidal 
THSI nuclei under consideration have such a distorted shape and low 
densities that they may probe the nuclear energy density functional in a 
new regime for which the extrapolation from the normal nuclear matter in 
the CDFT theory may not be adequate.  The possibility of using toroidal 
nuclei to probe the nuclear density functional at lower density may add 
an interesting dimension to the study of toroidal high-spin isomers.

\section{Summary and Conclusions}

In conclusion, the excitation function of the 7$\alpha$ channel in
deep-inelastic collisions of $^{28}$Si on $^{12}$C reveals resonance 
structures at high excitation energies. The features of these  structures 
appear to coincide with those  predicted by a number of theoretical
calculations in which the toroidal shell effects stabilize  the nucleus in a
diabatic $I$=0 state against major radius variations while spin-aligning 
particle-hole expectations lead to many high-spin toroidal isomers.  From 
the theoretical perspective, if the present results are confirmed by further 
studies, a very large number of diabatic and adiabatic toroidal high-spin 
isomers in a very large light-mass region may be opened up for future 
investigations.

Finally, we note that recent experimental and theoretical works provide 
indications that clustering effects are important in the collisions of 
alpha conjugate nuclei~\cite{schmidt17,schuetrumpf17}. The latter work, 
published as this paper was being prepared for submission provides 
theoretical evidence that toroidal alpha substructures can be quite 
commonly produced in such collisions. We strongly encourage further 
experimental work on collisions of light alpha-conjugate systems, both 
for the production of exotic clustered states and for the investigation 
of the dynamical evolution during such collisions. A higher granularity 
detector system and addition of gamma ray detection could offer significant 
improvements for such studies. 

\section{Acknowledgements}
This work was supported by the 
United States Department of Energy under Grant \# DE-FG03-93ER40773 and under 
grant \# DE-AC05-00OR22725 with UT-Battelle, LLC (Oak Ridge National 
Laboratory) and by The Robert A. Welch Foundation under Grant \# A0330. 
Partial support by the National Natural Science Foundation of China under 
Contracts No. 11421505, No. 11335002, No. 11621131001, and No. 11305239 and 
the Youth Innovation Promotion Association CAS (No. 2017309) are 
acknowledged.  Travel support for C. Y. Wong under the CUSTIPEN Program is 
thankfully acknowledged.  We appreciate useful communications from A. Ono, 
J. Maruhn, T. Ichikawa and S. Umar.  We also greatly appreciate the efforts 
of the staff of the TAMU Cyclotron Institute.

\end{document}